\documentclass{article}
\usepackage{arxiv}

\usepackage[utf8]{inputenc} % allow utf-8 input
\usepackage[T1]{fontenc}    % use 8-bit T1 fonts
\usepackage{hyperref}       % hyperlinks
\usepackage{url}            % simple URL typesetting
\usepackage{booktabs}       % professional-quality tables
\usepackage{amsfonts}       % blackboard math symbols
\usepackage{nicefrac}       % compact symbols for 1/2, etc.
\usepackage{microtype}      % microtypography
\usepackage{amsmath}
\usepackage{amssymb}
\usepackage{xcolor}
\usepackage{graphicx}
\usepackage{epstopdf}
\usepackage{epsfig}
\usepackage{caption}
\usepackage{natbib}
\usepackage{doi}

\begin{document}

\title{Linking electrical and thermal conductivity through cross-property inclusion modelling}

\date{\today}

\author{P. A. Cilli$^*$\\
	Department of Earth Sciences\\
	University of Oxford\\
	South Parks Road\\
	Oxford, OX1 3AN, UK\\
	$^*$\texttt{phillip.cilli@earth.ox.ac.uk} \\
	\And
	M. Chapman\\
	School of GeoSciences\\
	The University of Edinburgh\\
	James Hutton Rd, King's Buildings\\
	Edinburgh, EH9 3FE, UK\\
}

\renewcommand{\shorttitle}{Thermal-electrical modelling}

%%% Add PDF metadata to help others organize their library
%%% Once the PDF is generated, you can check the metadata with
%%% $ pdfinfo template.pdf
\hypersetup{
pdftitle={Linking electrical and thermal conductivity through cross-property inclusion modelling},
pdfauthor={P. A. Cilli, M. Chapman},
pdfkeywords={Cross-property modelling, Thermal conductivity, Electrical conductivity, DEM, Inclusion modelling},
pdfsubject={cond-mat.mtrl-sci}
}

\maketitle

\begin{abstract}
We derive a new cross-property differential effective medium scheme for a composite material's thermal conductivity as a function of its electrical conductivity and vice versa. Our scheme assumes that one phase is embedded in the other as inclusions. The relations are independent of inclusion volume fraction, but depend on the aspect ratio of the inclusions. We show that the method successfully models published laboratory measurements on a copper-graphite composite, with the inferred aspect ratio matching the physical shape of the inclusions. This work complements earlier results on elastic-electrical cross-property differential effective medium modelling, and has the potential to be extended for different cross-property relationships.
\end{abstract}

keywords={Cross-property modelling, Thermal conductivity, Electrical conductivity, DEM, Inclusion modelling}

\section{Introduction}

The mathematical relations linking a composite's various effective physical properties, known as cross-property relations, can be useful when one of the composite's properties is easier to measure than that which we would like to quantify \citep{Gibiansky1996,Carcione2007}. Rigorous bounds have been developed in the form of inequalities \citep{Torquato2002,Milton2002,Berryman1988,Gibiansky1996,Mavko2013,Carcione2007} which specify the physically permissible range for a composite's effective physical properties. Exact cross-property relations for composites have also been developed \citep{Pabst2015,Levin1967,Pabst2006,Sevostianov2002,Sevostianov2006,Pabst2017}, but there has been less attention given specifically to the development of a simple cross-property model derived from first principles which has a single, physically intuitive model parameter, no inclusion volume fraction terms, and the potential to generalise to model numerous cross-property relations.

\citet{Cilli2021} derived an electrical-elastic cross-property model using the differential effective medium (DEM) approximation (e.g., \cite{Berryman1992,Mendelson1982,Torquato2002}), which was shown to accurately estimate the elastic properties of brine-saturated clean sandstone cores from their measured electrical conductivity, and vice versa. This electrical-elastic model is derived from first principles, has no inclusion volume fraction terms, yet is correct in the high and low inclusion volume fraction (porosity) limits.

As a composite's effective thermal conductivity, electrical permittivity, magnetic permeability, and diffusion constant can be estimated by inclusion models \citep{Choy2016} such as the DEM approximation, \citet{Cilli2021} conjectured that the cross-property DEM approximation may also relate any of these physical properties to one another in a composite, resulting in a generalised cross-property relation which has no inclusion volume fraction terms.

In this paper we extend the electrical-elastic cross-property DEM approximation of \citet{Cilli2021} to model the forward and inverse relationships between a composite's thermal and electrical conductivities. Like the electrical-elastic cross-property DEM model, the models we present here are derived from first principles and have no inclusion volume fraction terms. We find these models accurately fit the thermal-electrical measurements of \citet{Mazloum2016} made on copper-graphite composites and that there is good agreement between the best-fitting model parameter, inclusion aspect ratio, and the experimentally measured aspect ratio of the graphite flake inclusions.

To begin, we derive the thermal-electrical cross-property DEM model before presenting our modelling method, including an overview of the public measurements which are modelled. Following this, we present then discuss our modelling results.

\section{Theory}

\subsection{Electrical and Thermal Modelling} \label{Subsec:Electrical}

In the case of a two-phase composite containing randomly oriented spheroidal inclusions, the electrical DEM model of \citet{Mendelson1982} can be expressed as \citep{Torquato2002,Cilli2021}

\begin{equation}
\frac{d\sigma^*}{d\phi}=\frac{\left( \sigma_2-\sigma^* \right) \bar{m}}{\left(1-\phi\right)}\,, \label{eqn:ElecDEM}
\end{equation}

where $\sigma^*$ is the effective electrical conductivity, $\sigma_2$ is the conductivity of the inclusion phase, $\phi$ is the inclusion volume fraction, and

\begin{equation}\label{eqn:mbar}
\bar{m} = \frac{1}{3}\sigma^*\left[ \frac{4}{\sigma^*+\sigma_2+L\left(\sigma^*-\sigma_2\right)} + \frac{1}{\sigma^*-L\left(\sigma^*-\sigma_2\right)}\right]\,.
\end{equation}

Parameter $L$ (e.g., \citet{Osborn1945}) is the principal depolarisation factor for a spheroid, and is a function of its aspect ratio, $\alpha$. Parameter $L$ relates the scalar component of an external potential field along the spheroid's axis of unique length (or any axis in the case of a sphere) to the spheroid's dipole moment along the same axis. \citet{Osborn1945} presents formulae to evaluate $L$ for an ellipsoid of arbitrary aspect ratio(s). Equation \ref{eqn:ElecDEM} is solved with boundary condition $\sigma^*(\phi = 0) = \sigma_1$, where $\sigma_1$ is the background material's electrical conductivity.

This composite's thermal conductivity, electrical permittivity, magnetic permeability, and diffusion constant can also be modelled using inclusion models \citep{Choy2016} such as equation \ref{eqn:ElecDEM}, as these physical properties are all governed by the Laplace equation when the source of the external potential field is located at infinity, which is assumed in the typical inclusion model derivation (e.g., \citet{Eshelby1957}). As such, we can reinterpret the DEM model of \citet{Mendelson1982} to estimate a composite's effective thermal conductivity by replacing electrical conductivity terms $\sigma$ with thermal conductivity terms $\kappa$ in equations \ref{eqn:ElecDEM} and \ref{eqn:mbar}, yielding

\begin{equation}
\frac{d\kappa^*}{d\phi}=\frac{\left( \kappa_2-\kappa^* \right) \bar{t}}{\left(1-\phi\right)}\, \label{eqn:ThermDEM}
\end{equation}

where

\begin{equation}\label{eqn:tbar}
\bar{t} = \frac{1}{3}\kappa^*\left[ \frac{4}{\kappa^*+\kappa_2+L\left(\kappa^*-\kappa_2\right)} + \frac{1}{\kappa^*-L\left(\kappa^*-\kappa_2\right)}\right]\,.
\end{equation}

Analogous to equation \ref{eqn:ElecDEM}, equation \ref{eqn:ThermDEM} is solved with boundary condition $\kappa^*(\phi = 0) = \kappa_1$, where $\kappa_1$ is the background material's thermal conductivity.

\subsection{Thermal-Electrical Modelling}

\citet{Cilli2021} proposed an electrical-elastic cross-property model which was derived by applying the chain rule to an electrical and an elastic DEM model, rendering inclusion volume fraction a dummy variable and removing it from the resultant equations in the process. Here we apply the same approach to thermal-electrical modelling. We apply the chain rule to equations \ref{eqn:ElecDEM} and \ref{eqn:ThermDEM} to obtain the thermal-electrical cross-property DEM model

\begin{equation}
\frac{d\kappa^*}{d\sigma^*}=\frac{\left(\kappa_2-\kappa^*\right)}{\left( \sigma_2-\sigma^*\right)} \frac{\bar{t}}{\bar{m}}\,. \label{dkds}
\end{equation}

The inverse of equation \ref{dkds} is its reciprocal. However, it is also a forward-model in its own right when the chain rule is applied to estimate electrical conductivity from thermal conductivity, yielding

\begin{equation}
\frac{d\sigma^*}{d\kappa^*}=\frac{\left(\sigma_2-\sigma^*\right)}{\left(\kappa_2-\kappa^*\right)} \frac{\bar{m}}{\bar{t}}\,. \label{dsdk}
\end{equation}

Equation \ref{dkds} is integrated with boundary condition $\kappa^*\left(\sigma^*=\sigma_1 \right) = \kappa_1$, while equation \ref{dsdk} is integrated with boundary condition $\sigma^*\left(\kappa^*=\kappa_1 \right) = \sigma_1$.

In principle, the mathematical form of equations \ref{dkds} and \ref{dsdk} generalises to relate any two of a composite's electrical conductivity, thermal conductivity, electrical permittivity, magnetic permeability, and diffusion constant to one another without any inclusion volume fraction terms.

In the cases of spherical ($S$) inclusions ($\alpha=1$, $L=1/3$), disk-shaped ($D$) inclusions ($\alpha=0$, $L=1$), and needle-shaped ($N$) inclusions ($\alpha=\infty$, $L=0$), equation \ref{dkds} simplifies to

\begin{align}
\frac{d\kappa^*}{d\sigma^*}\Biggr|_\textrm{S}&=\frac{\kappa^*}{\sigma^*}\frac{\left(\kappa_2-\kappa^*\right)}{\left(\sigma_2-\sigma^*\right)}\frac{\left(2\sigma^*+\sigma_2\right)}{\left(2\kappa^*+\kappa_2\right)}\,; \label{dkdsSphere}\\
\frac{d\kappa^*}{d\sigma^*}\Biggr|_\textrm{D}&=\frac{\sigma_2}{\kappa_2}\frac{\left(\kappa_2-\kappa^*\right)}{\left(\sigma_2-\sigma^*\right)}\frac{\left(2\kappa_2+\kappa^*\right)}{\left(2\sigma_2+\sigma^*\right)}\,; \label{dkdsDisk}\\
\frac{d\kappa^*}{d\sigma^*}\Biggr|_\textrm{N}&=\frac{\left(\sigma_2+\sigma^*\right)}{\left(\kappa_2+\kappa^*\right)}\frac{\left(\kappa_2-\kappa^*\right)}{\left(\sigma_2-\sigma^*\right)}\frac{\left(\kappa_2+5\kappa^*\right)}{\left(\sigma_2+5\sigma^*\right)}\,. \label{dkdsNeedle}
\end{align}

The analogous special cases of equation \ref{dsdk} for spherical, disk-, and needle-shaped inclusions are the reciprocals of equations \ref{dkdsSphere} to \ref{dkdsNeedle} respectively.

\section{Method}

To test the accuracy of equations \ref{dkds} and \ref{dsdk}, we modelled the thermal-electrical laboratory measurements of \citet{Mazloum2016}. The measurements of \citet{Mazloum2016} were made on 5 samples of copper containing a dispersion of randomly oriented graphite flakes at different volume fractions. The volume fraction of graphite flakes for each sample were 0, 0.1, 0.3, 0.4, and 0.5, and the oblate spheroidal flakes had an experimentally measured aspect ratio of 0.1. The flakes themselves possessed transverse isotropy in both their thermal and electrical conductivities.

Following \citet{Mazloum2016}, the thermal and electrical conductivities used in modelling were taken from \citet{Kovavcik2011} and \citet{Kovavcik1996} respectively. The thermal and electrical conductivities along a graphite flake's unique (short) axis were 274 W/m K and $0.59 \times 10^{-8}$ $1/\Omega$ m respectively, and were 10 W/m K and $2.26 \times 10^{-8}$ $1/\Omega$ m respectively in the plane formed by the flake's two equal (long) axes. The copper background was isotropic, with thermal and electrical conductivities taken to be 348.6 W/m K and $58.8 \times 10^{-8}$ $1/\Omega$ m respectively.

Due to the anisotropy of the inclusion's conductivities, there were three unknown parameters in equations \ref{dkds} and \ref{dsdk}: inclusion aspect ratio and the effective thermal and electrical conductivities of the inclusion phase. We assumed the effective inclusion conductivities were bounded by the values measured along the long and short axes of the graphite flakes. We also assumed the effective inclusion phase conductivities to be isotropic as the flakes were randomly oriented in the samples.

We solved for these three unknown parameters by inverting equation \ref{dkds} using the measurements of \citet{Mazloum2016} made on samples with inclusion volume fraction between 0 and 0.4, inclusive. We did not include the sample comprised of  50\% graphite in this model parametrisation as inclusion models such as cross-property DEM assume a dilute dispersion of inclusions. We solved for all three parameters simultaneously by minimising the $l_2$-norm of the misfits in measured and modelled thermal conductivity. The search for optimal aspect ratio was over the solution space $[0,\infty]$, ranging from disks to spheres to needles.

With the model parameterised, we forward modelled effective thermal and electrical conductivity trends using equations \ref{dkds} and \ref{dsdk} respectively for inclusion aspect ratios $\alpha \in \{0,10^{-2},10^{-1},1,\infty\}$. We also evaluated the thermal-electrical Hashin-Shtrikman bounds using stratagem of \citet{Carcione2007}, whereby we rearranged the electrical Hashin-Shtrikman bounds in terms of inclusion volume fraction then substituted these into the expressions for thermal Hashin-Shtrikman bounds, and vice versa.

\section{Results}

The effective inclusion thermal and electrical conductivities were solved to be 24.1 W/m K and $0.61 \times 10^{-8}$  $1/\Omega$ m respectively. Thus it seems the isotropic effective thermal and electrical conductivities of the randomly oriented anisotropic graphite inclusions were dominated by their properties in the directions of lower conductivities. For thermal conductivity, this direction is parallel to the spheroid's symmetry plane, while for electrical conductivity, this direction is perpendicular to its symmetry plane. The inclusion aspect ratio was solved to be 0.097, which shows good agreement with the experimentally measured aspect ratio of 0.1.

\begin{figure}
\centering
\includegraphics[width=\linewidth]{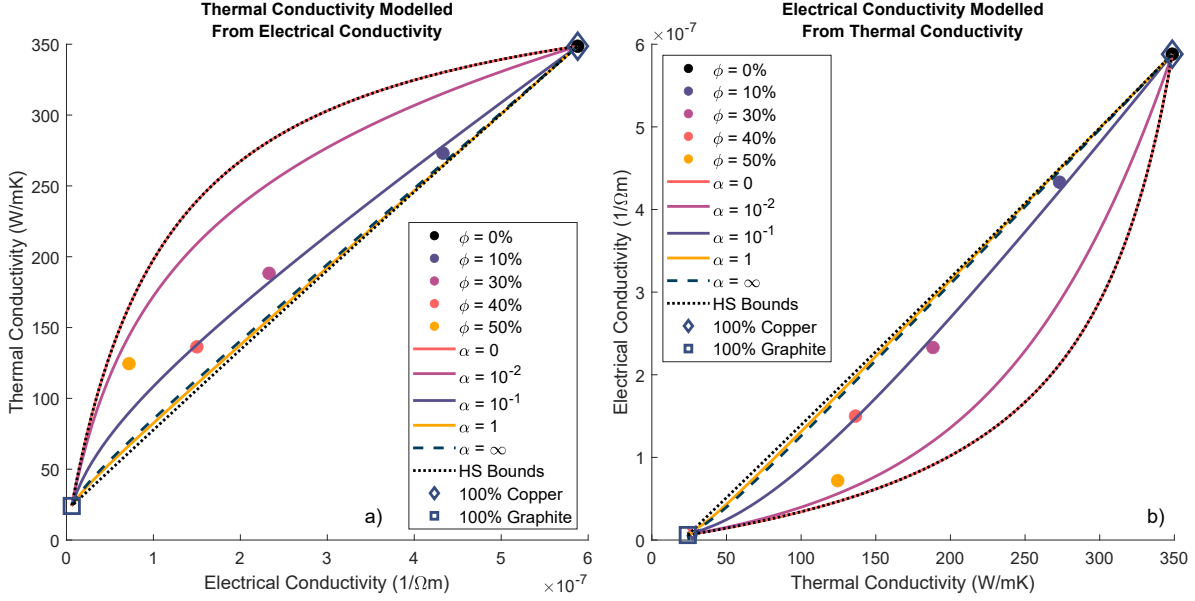}
\caption{Thermal-electrical cross-property modelling using cross-property DEM. Forward modelled a) thermal and b) electrical conductivity trends are generated by numerically solving equations \ref{dkds} and \ref{dsdk} respectively. Trends of constant inclusion aspect ratio $\alpha$ are seen to lie on or within the Hashin-Shtrikman bounds. All curves converge at the high (square) and low (diamond) inclusion volume fraction limits, where the model is correct. The measured data (circles) with inclusion volume fraction up to and including 0.4 all lie near the modelled curve for $\alpha=0.1$, which is the experimentally measured inclusion aspect ratio.}
\label{fig:model}
\end{figure}

Figure \ref{fig:model} shows forward modelled thermal and electrical conductivity trends using equations \ref{dkds} and \ref{dsdk} with the solved effective inclusion conductivities for various aspect ratios. The curves for $\alpha=1,0,\infty$ were created by solving equations \ref{dkdsSphere} to \ref{dkdsNeedle} respectively, and their reciprocals.

The model is seen to be correct in the high and low inclusion volume fraction limit. The modelled trend for disk-shaped inclusions seems to fall on the lower Hashin-Shtrikman bound, which may be expected as single-property DEM approximations can equal the lower Hashin-Shtrikman bound in the case of disk-shaped inclusions (e.g., \citet{Norris1985}).

Figures \ref{fig:model}a and \ref{fig:model}b are evidently the mirror-images of one another about the $x=y$ line. This is expected as equation \ref{dkds} is both the reciprocal and inverse of \ref{dsdk}. Nevertheless, all curves shown were calculated independently by numerically solving equation \ref{dkds}, \ref{dsdk}, or their special cases, equations \ref{dkdsSphere} to \ref{dkdsNeedle} and their reciprocals. 

\section{Discussion}

In the case of porous rocks, the inclusion aspect ratio parameter does not express the literal shape of pores (e.g., \citet{Fournier2011,Fournier2014,Fournier2018,Cilli2020,Cilli2021}). This is due to the presence of non-ellipsoidal pore shapes and the violation of the inclusion modelling assumption that inclusions are non-interacting. However, the composites of \citet{Mazloum2016} approximately follow the isotropic inclusion model's assumption of being a dilute dispersion of randomly oriented spheroids in a homogeneous background. By inferring an inclusion aspect ratio which agrees with the experimentally measured aspect ratio, we verify the accuracy of the cross-property DEM model in the case of its assumptions being approximately honoured.

Here we present an isotropic inclusion model where inclusions are assumed to be randomly oriented in the composite. This assumption appears to be valid when modelling the composites of \citet{Mazloum2016} as there was a good match between inferred and measured inclusion aspect ratios. However, \cite{Singh2020} showed using digital rock physics modelling that the aspect ratio parameter for the isotropic elastic DEM approximation of \citet{Berryman1992} depends upon the direction of sampling when applied to anisotropic rocks. Similarly, we would expect that the isotropic cross-property DEM model presented here would, in general, be inappropriate for modelling the properties of anisotropic rocks and composites, and an anisotropic cross-property model would be required in general.

The accuracy of the electrical-elastic cross-property DEM model of \citet{Cilli2021} without any inclusion volume fraction (porosity) terms suggests a clean sandstone's effective electrical and elastic properties vary with porosity in the same way. Similarly, we can infer from the observed accuracy of equations \ref{dkds} and \ref{dsdk} that the effective electrical and thermal conductivities of the modelled composite also depend on its inclusion volume fraction in the same way.

\citet{Cilli2021} speculated that their electrical-elastic cross-property DEM model could be extended to link other physical properties which can be modelled by inclusion modelling in their own right. Here we have shown this is true in the case of linking a composite's thermal and electrical conductivities.

Figure \ref{fig:model} shows both forward and inverse modelling using an inclusion aspect ratio of 0.1 accurately fits samples with inclusion volume fraction up to and including 0.4. Similarly, \citet{Cilli2021} showed forward and inverse modelling the electrical-elastic properties of clean sandstone cores using a single inclusion aspect ratio was accurate. What remains to be shown is whether inclusion aspect ratio is the same when linking three or more properties using multiple cross-property DEM relations on the same composite.

\section{Conclusions}

We have derived a cross-property model which relates a composite's thermal and electrical conductivities. This model is independent of inclusion volume fraction, depending only on inclusion shape. We have modelled published data and observed good agreement between the measured and inverted inclusion aspect ratios. This work builds on a similar electrical-elastic model, and we note its potential to extend to other cross-property relations.

\bibliographystyle{abbrvnat}
%\bibliography{FullBib}

\begin{thebibliography}{27}
\providecommand{\natexlab}[1]{#1}
\providecommand{\url}[1]{\texttt{#1}}
\expandafter\ifx\csname urlstyle\endcsname\relax
  \providecommand{\doi}[1]{doi: #1}\else
  \providecommand{\doi}{doi: \begingroup \urlstyle{rm}\Url}\fi

\bibitem[Berryman(1992)]{Berryman1992}
J.~G. Berryman.
\newblock Single-scattering approximations for coefficients in {Biot's}
  equations of poroelasticity.
\newblock \emph{The Journal of the Acoustical Society of America}, 91\penalty0
  (2):\penalty0 551--571, 1992.

\bibitem[Berryman and Milton(1988)]{Berryman1988}
J.~G. Berryman and G.~W. Milton.
\newblock Microgeometry of random composites and porous media.
\newblock \emph{Journal of Physics D: Applied Physics}, 21\penalty0
  (1):\penalty0 87, 1988.

\bibitem[Carcione et~al.(2007)Carcione, Ursin, and Nordskag]{Carcione2007}
J.~M. Carcione, B.~Ursin, and J.~I. Nordskag.
\newblock Cross-property relations between electrical conductivity and the
  seismic velocity of rocks.
\newblock \emph{Geophysics}, 72\penalty0 (5):\penalty0 E193--E204, 2007.

\bibitem[Choy(2016)]{Choy2016}
T.~Choy.
\newblock \emph{Effective Medium Theory: Principles and Applications}.
\newblock International Series of Monographs on Physics. Oxford University
  Press, 2016.
\newblock ISBN 9780198705093.

\bibitem[Cilli and Chapman(2020)]{Cilli2020}
P.~A. Cilli and M.~Chapman.
\newblock The power-law relation between inclusion aspect ratio and porosity:
  Implications for electrical and elastic modeling.
\newblock \emph{Journal of Geophysical Research: Solid Earth}, 125\penalty0
  (5):\penalty0 1--25, 2020.

\bibitem[Cilli and Chapman(2021)]{Cilli2021}
P.~A. Cilli and M.~Chapman.
\newblock Linking elastic and electrical properties of rocks using
  cross-property {DEM}.
\newblock Accepted to \emph{Geophysical Journal International}, 02 2021. DOI:10.1093/gji/ggab046.

\bibitem[Eshelby(1957)]{Eshelby1957}
J.~D. Eshelby.
\newblock The determination of the elastic field of an ellipsoidal inclusion,
  and related problems.
\newblock \emph{Proceedings of the Royal Society of London A}, 241\penalty0
  (1226):\penalty0 376--396, 1957.

\bibitem[Fournier et~al.(2011)Fournier, L{\'e}onide, Biscarrat, Gallois,
  Borgomano, and Foubert]{Fournier2011}
F.~Fournier, P.~L{\'e}onide, K.~Biscarrat, A.~Gallois, J.~Borgomano, and
  A.~Foubert.
\newblock Elastic properties of microporous cemented grainstones.
\newblock \emph{Geophysics}, 76\penalty0 (6):\penalty0 E211--E226, 2011.

\bibitem[Fournier et~al.(2014)Fournier, L{\'e}onide, Kleipool, Toullec,
  Reijmer, Borgomano, Klootwijk, and Van Der~Molen]{Fournier2014}
F.~Fournier, P.~L{\'e}onide, L.~Kleipool, R.~Toullec, J.~J. Reijmer,
  J.~Borgomano, T.~Klootwijk, and J.~Van Der~Molen.
\newblock Pore space evolution and elastic properties of platform carbonates
  ({Urgonian} limestone, {Barremian}--{Aptian}, {SE} {France}).
\newblock \emph{Sedimentary Geology}, 308:\penalty0 1--17, 2014.

\bibitem[Fournier et~al.(2018)Fournier, Pellerin, Villeneuve, Teillet, Hong,
  Poli, Borgomano, L{\'e}onide, and Hairabian]{Fournier2018}
F.~Fournier, M.~Pellerin, Q.~Villeneuve, T.~Teillet, F.~Hong, E.~Poli,
  J.~Borgomano, P.~L{\'e}onide, and A.~Hairabian.
\newblock The equivalent pore aspect ratio as a tool for pore type prediction
  in carbonate reservoirs.
\newblock \emph{AAPG Bulletin}, 102\penalty0 (7):\penalty0 1343--1377, 2018.

\bibitem[Gibiansky and Torquato(1996)]{Gibiansky1996}
L.~Gibiansky and S.~Torquato.
\newblock Bounds on the effective moduli of cracked materials.
\newblock \emph{Journal of the Mechanics and Physics of Solids}, 44\penalty0
  (2):\penalty0 233--242, 1996.

\bibitem[Kov{\'a}{\v{c}}ik and Bielekt(1996)]{Kovavcik1996}
J.~Kov{\'a}{\v{c}}ik and J.~Bielekt.
\newblock Electrical conductivity of {C}u/graphite composite material, as a
  function of structural characteristics.
\newblock \emph{Scripta materialia}, 35\penalty0 (2):\penalty0 151--156, 1996.

\bibitem[Kov{\'a}{\v{c}}ik and Emmer(2011)]{Kovavcik2011}
J.~Kov{\'a}{\v{c}}ik and {\v{S}}.~Emmer.
\newblock Thermal expansion of {C}u/graphite composites: effect of copper
  coating.
\newblock \emph{Kovove Mater}, 49:\penalty0 411--416, 2011.

\bibitem[Levin(1967)]{Levin1967}
V.~M. Levin.
\newblock Thermal expansion coefficient of heterogeneous materials.
\newblock \emph{Mechanics of Solids}, 2:\penalty0 58--61, 1967.

\bibitem[Mavko and Saxena(2013)]{Mavko2013}
G.~Mavko and N.~Saxena.
\newblock Embedded-bound method for estimating the change in bulk modulus under
  either fluid or solid substitution.
\newblock \emph{Geophysics}, 78\penalty0 (5):\penalty0 L87--L99, 2013.

\bibitem[Mazloum et~al.(2016)Mazloum, Kov{\'a}{\v{c}}ik, Emmer, and
  Sevostianov]{Mazloum2016}
A.~Mazloum, J.~Kov{\'a}{\v{c}}ik, {\v{S}}.~Emmer, and I.~Sevostianov.
\newblock Copper--graphite composites: thermal expansion, thermal and
  electrical conductivities, and cross-property connections.
\newblock \emph{Journal of Materials Science}, 51\penalty0 (17):\penalty0
  7977--7990, 2016.

\bibitem[Mendelson and Cohen(1982)]{Mendelson1982}
K.~S. Mendelson and M.~H. Cohen.
\newblock The effect of grain anisotropy on the electrical properties of
  sedimentary rocks.
\newblock \emph{Geophysics}, 47\penalty0 (2):\penalty0 257--263, 1982.

\bibitem[Milton(2002)]{Milton2002}
G.~W. Milton.
\newblock \emph{The Theory of Composites}.
\newblock Cambridge Monographs on Applied and Computational Mathematics.
  Cambridge University Press, 2002.

\bibitem[Norris(1985)]{Norris1985}
A.~N. Norris.
\newblock A differential scheme for the effective moduli of composites.
\newblock \emph{Mechanics of Materials}, 4\penalty0 (1):\penalty0 1--16, 1985.

\bibitem[Osborn(1945)]{Osborn1945}
J.~A. Osborn.
\newblock Demagnetizing factors of the general ellipsoid.
\newblock \emph{Physical Review}, 76\penalty0 (11-12):\penalty0 351--357, 1945.

\bibitem[Pabst and Gregorov{\'a}(2006)]{Pabst2006}
W.~Pabst and E.~Gregorov{\'a}.
\newblock Cross-property relations between elastic and thermal properties of
  porous ceramics.
\newblock In \emph{Advances in Science and Technology}, volume~45, pages
  107--112. Trans Tech Publ, 2006.

\bibitem[Pabst and Gregorov{\'a}(2015)]{Pabst2015}
W.~Pabst and E.~Gregorov{\'a}.
\newblock Elastic and thermal properties of porous materials--rigorous bounds
  and cross-property relations (critical assessment 18).
\newblock \emph{Materials Science and Technology}, 31:\penalty0 1801--1808,
  2015.

\bibitem[Pabst and Gregorov{\'a}(2017)]{Pabst2017}
W.~Pabst and E.~Gregorov{\'a}.
\newblock A generalized cross-property relation between the elastic moduli and
  conductivity of isotropic porous materials with spheroidal pores.
\newblock \emph{Ceram. Silik.}, 61\penalty0 (1):\penalty0 74--80, 2017.

\bibitem[Sevostianov and Kachanov(2002)]{Sevostianov2002}
I.~Sevostianov and M.~Kachanov.
\newblock Explicit cross-property correlations for anisotropic two-phase
  composite materials.
\newblock \emph{Journal of the Mechanics and Physics of Solids}, 50\penalty0
  (2):\penalty0 253--282, 2002.

\bibitem[Sevostianov et~al.(2006)Sevostianov, Kov{\'a}{\v{c}}ik, and
  Siman{\v{c}}{\'\i}k]{Sevostianov2006}
I.~Sevostianov, J.~Kov{\'a}{\v{c}}ik, and F.~Siman{\v{c}}{\'\i}k.
\newblock Elastic and electric properties of closed-cell aluminum foams:
  cross-property connection.
\newblock \emph{Materials Science and Engineering: A}, 420\penalty0
  (1-2):\penalty0 87--99, 2006.

\bibitem[Singh et~al.(2020)Singh, Cilli, Hosa, and Main]{Singh2020}
J.~Singh, P.~Cilli, A.~Hosa, and I.~Main.
\newblock Digital rock physics in four dimensions: simulating cementation and
  its effect on seismic velocity.
\newblock \emph{Geophysical Journal International}, 222\penalty0 (3):\penalty0
  1606--1619, 2020.

\bibitem[Torquato and Haslach~Jr(2002)]{Torquato2002}
S.~Torquato and H.~Haslach~Jr.
\newblock Random heterogeneous materials: microstructure and macroscopic
  properties.
\newblock \emph{Appl. Mech. Rev.}, 55\penalty0 (4):\penalty0 B62--B63, 2002.

\end{thebibliography}

\end{document}